\documentstyle[preprint,floats,aps,epsf,epsfig,prb]{revtex}
\title{Thickness-Dependence of the Coercive Field in Ferroelectrics}
\author{P. Chandra,$^1$ M. Dawber,$^2$ P.B. Littlewood$^3$ and J.F. Scott$^2$}
\address{$^{1}$NEC, 4 Independence Way, Princeton, NJ 08540, U.S.A.}
\address{$^{2}$Symetrix Centre for Ferroics, Earth Sciences
Department, Downing Street, University of Cambridge, Cambridge U.K.
CB2 3EQ}
\address{$^{3}$Cavendish Laboratory, University of Cambridge,
Madingley Road, Cambridge U.K. 
CB3 OHE} 

\newcommand \ltdash{\raise-1.8pt\hbox{$\scriptscriptstyle |$}}

\newcommand \bea {\begin{eqnarray} }
\newcommand \eea {\end{eqnarray}}

\newlength{\bxwidth}\bxwidth=0.8\textwidth

\newcommand\prm[2]{$ $\vskip 2.4 truein
\centerline{\epsfig{file=#1,width=\bxwidth} }\vskip 0.5truein
\centerline{{\bf Fig.} #2}}
\newcommand\prk[2]{$ $\vskip 2.4 truein\vskip -2 truein
\centerline {\epsfig{file=#1,width=\bxwidth}} \vskip 0.5truein
\centerline{{\bf Fig.} #2}}
\begin{document}
\maketitle
\begin{abstract}
For forty years researchers on ferroelectric switching have used the 
Kay-Dunn theory to model the thickness-dependence of the coercive 
field; it works surprisingly well, despite the fact that it is 
based upon homogeneous nucleation and a small-field expansion, 
neither of which is realized in thin films. Here we 
demonstrate that this result can be obtained from a more general 
Kolmogorov-Avrami model of (inhomogeneous) nucleation and growth.  By 
including a correction to the switching field across the 
dielectric that includes Thomas-Fermi screening in the metal 
electrode, we show that our theory quantitatively describes the coercive 
fields versus thickness in several different families of 
ferroelectric (lead zirconate-titanate [PZT], potassium nitrate, and 
polyvinylidenefluoride [PVDF]) over a wide range of 
thickness (5 decades).  This agreement is particularly satisfying
in the case of PVDF, as it indicates that the switching kinetics
are domain-wall limited down to 1 nanometer and thus
require no new effects.
\end{abstract}
\vskip 0.2truein
\pacs{77.55.+f, 77.80.Fm, 68.15.+e, 85.50.+k}

\newpage

Demand for integrated microelectronics with increased densities and decreasing
voltage standards requires detailed characterization of the coercive field,
$E_c$, for the design and fabrication of
competitive, reliable ferroelectric
devices.  More specifically, recent measurements on nanometer 
ferroelectric films\cite{Bune98}
have revealed a weakness in our current understanding 
of the thickness-dependence
of $E_c$.  
Motivated by these experiments, in this Letter we present a new
treatment of the coercive field as a function of thickness, $E_c(d)$, that incorporates
both inhomogenous nucleation in a finite-size film and field penetration in
the electrodes.  The $E_c(d)$ that emerges
from our approach is in good
agreement with observed behavior 
for films that range from
100 microns to 1 nanometer.  Furthermore our approach indicates 
that the minimum
film thickness, determined by the magnitude of the 
depolarization field,\cite{Batra72} can be tuned by varying
the spontaneous polarization of the ferroelectric and the screening length of
the electrodes.  Other implications of our results for the processing windows
of next-generation FeRAMS and DRAMS are also discussed.

For the last forty years the Kay-Dunn scaling law\cite{Kay62} has been
successfully
used to 
describe the thickness-dependence of the coercive field in 
ferroelectric films ranging from 100 microns to 200 nanometers.\cite{Scott00} 
However the Kay-Dunn model is 
based on assumptions, particularly that of homogenous nucleation,
which are inappropriate for thin films.
As discussed in the original
Kay-Dunn paper,\cite{Kay62} the calculated energy barrier associated with such 
nucleation is several orders of magnitude 
larger than any experimentally determined
value; furthermore the presence of long-range elastic interactions
increases
this barrier still further.\cite{Littlewood86}  Imaging measurements confirm the
site-specific nature
of nucleation in thin ferroelectric films, indicating that new domains nucleate
at electrodes and at twin boundaries.\cite{Lajzerowicz81,Shur91,Houchmandzadeh92,Ganpule00}  
The good agreement between measured values and the Kay-Dunn treatment of
$E_c(d)$ is thus quite
fortuitous.

Here we show that the Kay-Dunn scaling result can be recovered from an adapted
Kolmogorov-Avrami model\cite{Ishibashi95} of inhomogeneous
nucleation 
in a confined geometry.
In this approach
the transformation of a sample from  paraelectric to ferroelectric
is treated by considering the nucleation and growth of a 
single domain.  The coercive field 
is determined by the condition that
\begin{equation}
P(E,f) \left \vert_{E = E_c} = \frac{1}{2} \right.
\end{equation}
where $P(E,f)$ is the untransformed  sample fraction as a function of
applied field $E$ and frequency $f$.  For
inhomogeneous nucleation 
$P(E,f) \sim \exp -N(E) (v(E) \frac{2\pi}{f})^{D}$ 
where
$N(E)$ and $v(E)$ are the field-dependent 
number of nucleation sites per unit
area and domain growth velocity respectively and $D$ is
the dimension of the nucleus.
The factor $(v(E) \frac{2\pi}{f})^{D}$ enters the formula
for $P(E,f)$ as the characteristic volume of a growing nucleus
over the cycle time of the electric field.  However if the nuclei
elongate through the thickness of the film before colliding
with another domain, this factor should be replaced
by $(\kappa d)$ where $\kappa$ is the (small) transverse
area of the needle-shaped domain and $d$ is the film
thickness.  Such a morphology is supported by experimental
observation.\cite{Fatuzzo67}
Then 
we obtain
$P(E) \sim \exp -  N(E)  (\kappa d)$ where $N \sim E^\alpha$.
$\alpha$ is known to equal $3/2$ over a fairly
broad range of fields both experimentally\cite{Merz54}
and theoretically\cite{Stadler58} from switching kinetics; this then
leads to the desired result,
$E_c \sim d^{-2/3}$.
Here $0 < \kappa < 1$
indicates the size of the nucleating grain as a fraction of the
film thickness. We note that the scaling of $E_c$ with thickness follows directly from
the field-dependence of the number of nucleation sites per unit
area, $N(E)$; this result could be checked independently by
tuning
$N(E)$ and observing the predicted change in $E_c(d)$.

Recently it has been possible to measure the coercive field of
ferroelectric PVDF films with thicknesses less than 100 nanometers; 
on these length-scales significant deviation from the scaling
(cf. Figure 1) described
above is observed.\cite{Bune98}  In retrospect,
there were already hints of this new behavior in  
$E_c(d)$ for films between 100 and 200 
nanometers,\cite{Nagarajan99}
but the recent measurements down to 1 nanometer
are quite conclusive.\cite{Bune98}
The scaling treatment of $E_(d)$ assumes
perfect conducting plates, and here we have modified it
to include field-penetration in the electrodes.
We show that the latter effect becomes important
for small film thicknessses and is responsible for the observed
deviation from the expected scaling behavior (cf. Figure 1).  Furthermore we
demonstrate that, using these ideas, we can include a correction to the
measured coercive fields and recover the scaling result for films
ranging from 100 microns down to 1 nanometer (cf. Figure 2).

In an idealized ferroelectric capacitor, the plates are assumed to be
perfect conductors and charge resides on a plane
of negligible thickness at the electrode-ferroelectric interface,
compensating for the spontaneous polarization in the
ferroelectric
film.  Realistically this charge
distributed over a small but finite
length-scale in the metal, $\lambda$.  
The resulting electric field leads to an associated voltage 
drop in the metal electrodes.  
A compensating depolarization
potential, $\Phi_{dp}$,
must exist across the film 
to ensure that it is an equipotential 
in the absence of an externally applied voltage.
The associated field, $E_{dp} = \frac{\Phi_{dp}}{d}$,
assists in the switching
process so that the measured coercive field, $E_c^{meas}$, is
\begin{equation}
E_c^{meas} = E_c^{i} -  E_{dp}
\label{ec}
\end{equation}
where $E_c^{i}$ refers to a perfect capacitor with
negligible screening and is described by the Kay-Dunn scaling.

In order to determine $\Phi_{dp}$,
and its associated field,
we assume that the charge density takes the form
\begin{equation}
\rho(x) = 
\frac{Q}{\lambda} e^{ \left( - \frac{x}{\lambda}\right)}
\end{equation} 
and, 
furthermore we
assume symmetric capacitor plates; then, using the fact
that the potential in the electrodes is related to the induced
charge density,  we find 
\begin{equation}
\left \vert  \Phi_{dp} \right\vert = \left \vert \left(\frac{4
\pi}{\epsilon_m}\right) \frac{ 2\lambda Qd}{A}\right \vert
\end{equation}
where $\epsilon_m$ refers to the metallic dielectric constant,
and $Q = \frac{P_s A}{d}$ in a ferroelectric where $P_s$ is the
spontaneous polarization.
Here we estimate $\lambda$ using
the assumption of a slowly varying potential,\cite{Kittel76}
resulting in $\lambda^2 = \epsilon_m \lambda_{TF}^2$ where $\lambda_{TF}$
is the Thomas-Fermi screening length.   
Experimentally $P_s$ is observed to be independent of film
thickness in the nanometer PVDF samples studied\cite{Bune98},
indicating that $\Phi_{dp}$ is a constant; note that
in this case the voltage drop across the electrodes, $V_{electrodes}$,
compensates for the depolarization potential $\Phi_{dp}$.
However $E_{dp}$
scales inversely with $d$, so that it assumes increasing
importance in the expression for $E_c^{meas}$, (\ref{ec}),
with decreasing film thickness.  

Field-penetration of the electrodes has been studied in the
context of very thin ($\sim 30$ angstroms) film capacitors
by several authors, \cite{Mead61,Ku64,Simmons65,Black99}
and here we argue that it is relevant for thicker
films whose dielectric constant, $\epsilon_f$ is
large.
The effective capacitance of the metal-film-metal sandwich 
can be determined by contributions from two capacitors
in series\cite{Ku64}
\begin{equation}
\frac{1}{C} = \frac{1}{C_{film}} + \frac{1}{C_{electrodes}}
\end{equation}
where $C_{film}$ and $C_{electrodes}$ are the film
and the electrode capacitances respectively;
the first term is geometric and the second provides
an upper bound to the capacitance with decreasing film thickness.
For a dielectric film, the voltage drop is
$V_{film} =  \frac{4\pi}{\epsilon_{f}} \left( \frac{Qd}{A}\right)d$,
so that the fraction of the total potential difference, $V = V_{film}
+ V_{electrode}$, across
the electrodes is
\begin{equation}
\frac{V_{electrodes}}{V} 
= \frac{2\lambda \left( \frac{\epsilon_f}{\epsilon_m}\right)}
{ d + 2\lambda 
\left(\frac{\epsilon_f}{\epsilon_m}\right)}
\end{equation}
which can be substantial for large $\epsilon_f$ and small $d$.
Therefore we note that field-penetration effects will be important
for both (non-switching) high-dielectric DRAM capacitors and in 
switching FeRAMS.

Let us now turn to a more quantitative analysis.
In Figure 1, we display the measured coercive field data\cite{Bune98} for
PVDF and the Kay-Dunn scaling prediction.  We note that there is
deviation from the scaling behavior at approximately 100 nanometers.
Next, using (\ref{ec}), we plot the coercive field including
the effect of field-penetration.  We have used the value
of $\epsilon_f = 14$
for PVDF from experiment,\cite{Furukawa81}
and used  
$\epsilon_m = 1$ and $\lambda_{TF} = 0.45$ angstroms for the aluminum
electrodes.\cite{Kittel76}
For thicknesses greater than 100 nm, the modified and the scaling
curves are identical, indicating that the depolarization contribution
is not important in this regime.  However
the modified $E_c$ continues to display a good fit to the data for
nanometer
thicknesses, particularly as it
deviates from the scaling curve.   We note that the observed rollover
in $E_c(d)$ will correspond to the thickness where the depolarization
and the idealized coercive fields are comparable; here we expect a
polarization
instability\cite{Batra72}
that will determine the minimum allowed film thickness, $d_{min}$.
We note that our results suggest that this instability
can be tuned by varying the spontaneous polarization of
the ferroelectric and the screening length of the electrodes.
This is consistent with the results of Ghosez and Rabe\cite{Ghosez00}
who found that $d_{min}$ for a ferroelectric
film with perfect electrodes was less than that found with
semiconducting ones.\cite{Batra72}

Alternatively we can use (\ref{ec}) to add depolarization corrections
to the measured coercive field; the resulting idealized $E_c^i$
displays Kay-Dunn scaling for five decades in thicknesses ranging
from 100 microns to 1 nanometers (cf. Figure 2).  We present the
modified data for three different materials in the main part of
the figure; since $E_c = C \times d^{-2/3}$ where $C$
is material-specific, we expect universal behavior for 
$\log E_c - \log C$.  Indeed this is the case, as indicated
in the inset of Figure 2.  It is unusual to see universal scaling
over such a large number of decades.
The agreement between theory and experiment is particularly
satisfying for the case of PVDF, since it demonstrates
that the switching kinetics of films on nanometer scales are
domain-wall
limited and are qualitatively similar to those of their 100 micron
counterparts.  No unusual effects special to ultrathin films
are required; this has been a topic of discussion
in the literature.\cite{Fridkin00}

Our results have several implications for the design and fabrication
of ferroelectric devices.  Ideally FeRAMS should be designed at
film thicknesses where the measured coercive field is less than
the scaling value, but before it is equal to the depolarization
field; this gives a processing window for PVDF, the polymeric ferroelectric
used for this study, for 10 - 100 nm.  Furthermore the minimum
film thickness, associated with a polarization instability,
can be tuned by varying $P_s$ and $\lambda$.  Oxide electrodes,
known for reducing ferroelectric fatigue\cite{Scott00}, would
{\sl not} be optimal for very thin ($< 10$ nm) devices due to
large field-penetration in the capacitor plates.  Indeed the
optimization of very thin devices should emphasize the
screening properties of the electrodes, favoring
Pt or Au over $SrTiO_3$.  For the DRAM market, the primary objective
is maximizing total capacitance; our results indicate that it is
limited by the electrodes.  Thus use of larger dielectric materials
and thinner films will {\sl not} improve it significantly.  Instead
one should optimize the electrical resistivity of the ferroelectric
film and the screening properties of the electrodes.

In summary, in this Letter we have brought together the ideas of Kay and
Dunn,\cite{Kay62} Batra and Silverman,\cite{Batra72}
and Ghosez and Rabe\cite{Ghosez00} to develop a model for
the coercive field as a function of thickness.  For thin films,
we have shown the importance of field-penetration in the electrodes;
the resulting depolarization field is a function of the spontaneous
polarization of the ferroelectric and the Thomas-Fermi screening
length of the metal electrode.  Naturally when $E_{dp}$ becomes equal
to the idealized coercive field the polarization is unstable;\cite{Batra72}
however when it is comparable but less than $E_c^i$, it contributes
to the switching.  In Figure 1, with no adjustable parameters, we
display good agreement between our modified coercive field and
that measured in PVDF, particularly for thicknesses $<$ 100 nm
where there is deviation from scaling behavior.  In Figure 2
the measured data are corrected to accommodate field-penetration
in the electrodes; the result obeys Kay-Dunn scaling for
five decades in length-scale.  Our results indicate that the choice
of electrode is very important for optimizing performance for
very thin film devices.  Given increasing clock speeds, an
improved understanding of the coercive fields at finite
frequencies would also be of great interest for the development 
of future ferroelectric devices.

We are grateful to S. Bhattacharya, D.J. Jung, K. Rabe, A. Ruediger
and M.M.J. Treacy for discussions.

\newpage

\noindent{\bf Figure Captions}

\renewcommand{\labelenumi}{{\bf Fig.} \theenumi .}
\begin{enumerate}

\item 
Measured coercive field data$^1$ for PVDF,
the Kay-Dunn scaling and the current theory vs. thickness; as
explained in the text, the latter incorporates field-penetration effects via a
depolarization contribution to the field as a function of thickness.
\label{1}

\item 
{\bf (Main Figure)} The corrected log coercive field vs. log thickness
where the depolarization contribution has been subtracted from the
measured data {\bf (Inset)}
Normalized corrected 
log coercive field vs. thickness which indicates Kay-Dunn scaling
over five decades of lengths; here universal behavior is shown
for $\log E_c - \log C$ where $C$ is defined by the expression
$E_c = C\times d^{-2/3}$ and is material- and sample-specific.  
\label{2}

\end{enumerate}

\newpage


\prk{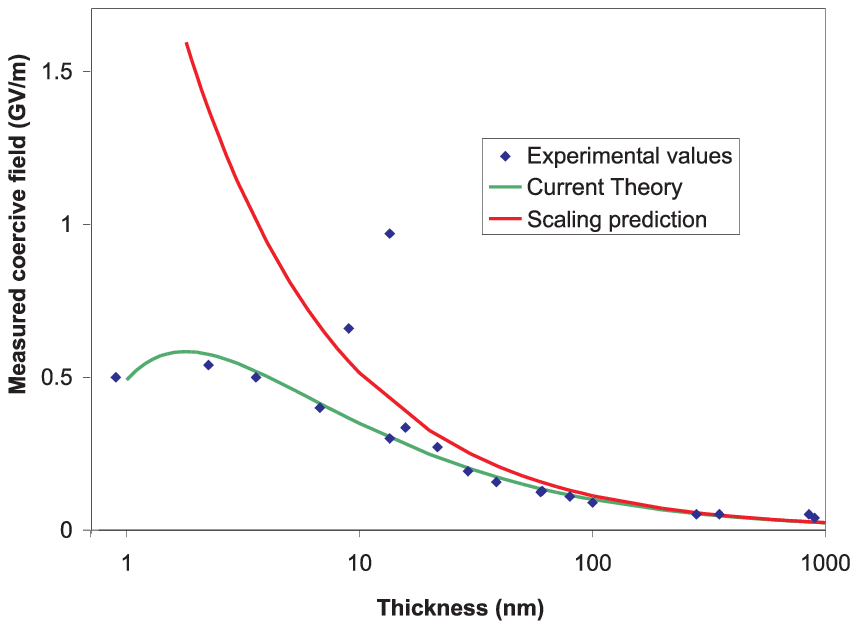}{1}



\prm{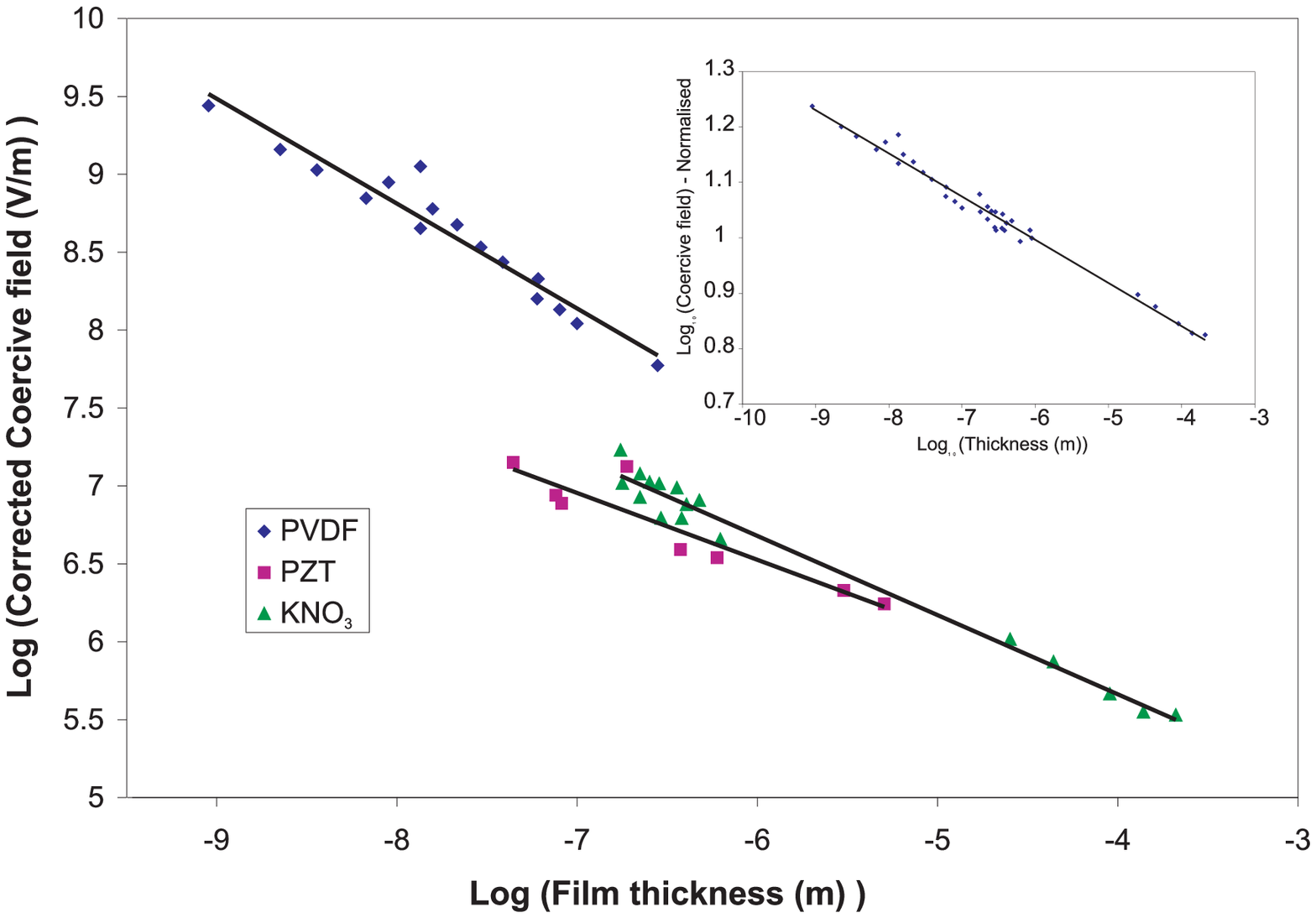}{2}

\end{document}